# A HIGH PERFORMANCE SCINTILLATOR ION BEAM MONITOR*

D. S. Levin†, University of Michigan, Ann Arbor, MI, USA
P. S. Friedman, Integrated Sensors LLC, Palm Beach Gardens, FL, USA
T. Ginter, Michigan State University, East Lansing, MI, USA
C. Ferretti, A. Kaipainen, N. Ristow, University of Michigan, Ann Arbor, MI, US

## Abstract

A high-performance Scintillator Ion Beam Monitor (SBM) provides diagnostics across an extremely wide range of isotopes, energies, and intensities employing a machine-vision camera with two novel scintillator materials moveable into/out of the beam without breaking vacuum. Scintillators are: 1) a semicrystalline polymer material (PM), film, 1-200 µm thick; and 2) a 100-400 µm thick opaque sheet consisting of a hybrid of an *inorganic-polymer* (HM) hybrid matrix. The SBM was demonstrated at the Facility for Rare Isotope Beams (FRIB, East Lansing, MI) providing real-time beam profile and rate analysis spanning more than five orders of magnitude *including visualization of single ion signals*. It may replace FRIB reference detectors: a phosphorescent beam viewer, a Faraday cup, a microchannel plate, and a silicon detector.

## INTRODUCTION

Ion beam laboratories such as the Facility for Rare Isotope Beams (FRIB) feature many experimental beam lines requiring extensive, and potentially very expensive beam tuning. For example, the 2023 FRIB operating budget is nearly $100M **[1]**, distributed over some 3100 active beam hours suggests an hourly beam operating cost on the order of $30K. Other facilities such as ANL, JLab, BNL/NASA require high performance scintillator detectors for charged particle counting over a wide dynamic range. High usage costs put a premium on technologies that enable the beam to be imaged, focused and otherwise tuned for research use with minimal time overhead and maximum versatility and spatial precision. The Scintillator Ion Beam Monitor system‡ (SBM) hardware/software beam detector and analysis suite is designed to meet these requirements. This article describes the SBM and recent experiment results.

The SBM hardware employs a machine-vision camera combined with a fast, large aperture lens and two proprietary thin scintillator materials. The scintillators mount on a multi-target cassette controlled by a robotic arm that allows them to be swapped or translated in/out of the beam without breaking vacuum. This detector is fabricated in a six-way stainless steel cross (6WC) composed of the three orthogonal transepts and can be configured directly into a high vacuum beamline. Figure 1 shows the 6WC mounted on a support stage for positioning in a beam, and a cross-sectional view showing the ladder or cassette of scintillator targets. A high resolution (megapixels), low-noise (2.4 photoelectrons RMS/ADC bin) CMOS sensor collects the scintillation light along the vertical branch line through a clear port window with a fast (*f/*0.9) lens. The field of view and depth of field are sufficient to image the entire target region.

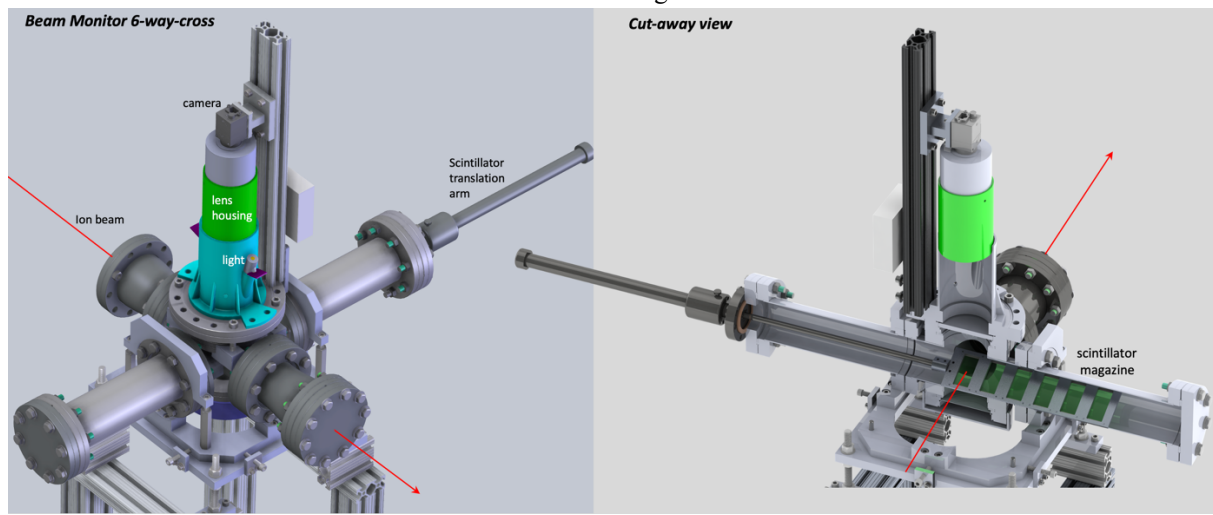

Figure 1: The six-way cross (6WC) mounted on a support stage (left); cut-away view showing the internal scintillator cassette (right).

### Scintillators

Two types of proprietary scintillators are employed in the SBM: (1) Polymer Material (PM) – a semicrystalline polymer developed as a resilient thin film primarily for packaging applications. It was discovered to be an intrinsic scintillator with superior physical properties and higher light-yield than most plastic scintillators based on polyvinyltoluene and/or polystyrene. Because PM is semicrystalline, it has a "hazy" appearance. The PM-scintillator is highly radiation damage resistant. It produces a much stronger photodetector signal in our 6WC than other plastic scintillators tested such as BC-400 [2]. Various scintillator thicknesses from 200 µm to 1 µm have been tested. These thin films are particularly attractive for transmissive applications such as external beam radiation therapy (EBRT) where it is necessary for the imaging target to remain in the beam during treatment; (2) Hybrid Material (HM) is an *inorganic-polymer* material, is non-hygroscopic, is radiation damage resistant, is available in both thin and large area sizes, and can generate up to order of magnitude stronger signals per unit thickness than CsI(Tl). These large signals are attributed primarily to the photon transport; photons do not internally reflect and are emitted readily from the surface. The HM photon yield itself per energy deposited might be similar to CsI(Tl), in the range of 48000-65000 photons/MeV [3] [4]. Being polycrystalline in nature, it is visually opaque and incapable of total internal reflection, thus resulting in (1) a higher percentage of photons escaping from the film surface, (2) reduced back surface reflection and more accurate beam imaging and dosimetry. Two types of HM ($HM_1$ and $HM_2$) are included for SBM use, differentiated by their dopants, light yields, and decay times in the µs versus ms range respectively.

* Work supported by SBIR Phase-II award from the DOE Office of Science to Integrated Sensors, LLC (Award No. DE-SC0019597).
† dslevin@umich.edu
‡ An Integrated Sensors patented product

## EXPERIMENTAL RESULTS

The SBM has been evaluated experimentally in various laboratories and with different radiation sources. Initial tests focused on scintillator performance using low-rate alpha and beta sources ($^{241}$Am, $^{90}$Sr). Beam tests at FRIB using a 3 MeV/n $^{86}$Kr beam in Sept. 2021, were followed at the Michigan Ion Beam Lab (MIBL) with high intensity 1-5 MeV protons, and at the Notre Dame Radiation Laboratory using fast, intense bunches of 8 MeV electrons. Selected results are presented here.

*Benchtop Scintillator Tests*

Initial test-bench experiments were conceived to evaluate HM and PM scintillator performance with respect to reference standards. The setup was arranged in a dark box using the 6WC components and a similar optical layout. The camera imaged 2 x 4 cm$^2$, 105-435 µm thick HM and 1.25 mm thick unpolished CsI(Tl), scintillator targets over a range of viewing angles. The radiation source was a 2.4 mCi $^{90}$Sr button that produced a 3 mm diameter beam of ~1 MeV β$^-$ particles. Images of this beam are shown in Figure 2 for a 0° viewing angle. The top panel shows the beam imaged on CsI(Tl), which may be compared to the same beam on a $HM_2$ target in the lower panel. The intense inner beam core corresponds to the 3 mm source bore. The significant feature of these tests is 1) the HM image is clean with well-defined beam since there are no internal reflections that distort or otherwise cause the beam image to bloom; 2) the relative signal strength of HM to CsI(Tl) scintillator, normalized to the material thickness: (units of ADC counts/mm) ranges from 5.4 to 12.2 (± 10%) for $HM_1$ & $HM_2$ respectively. Furthermore, we note that the tested *unpolished* CsI would yield more front surface photon emission than a polished crystal. Using the same experimental setup the signal from commercial BC-400 scintillator relative to PM type of similar thickness was compared. The results are shown in Figures 3 top and bottom respectively. The ADC distributions pertain to the hits within a fiducial signal region indicated by the red square frame. The PM image is much more filled in and the average ADC value is nearly 5 times larger. We note that this is not attributed to higher light yields as both materials likely have similar efficiency. Because PM is semi-crystalline and hazy, photons easily escape the planar surfaces towards the lens. The BC-400 is clear with smooth, specular surfaces; light internally reflects and escapes mostly out the edges where it is not collected by the lens.

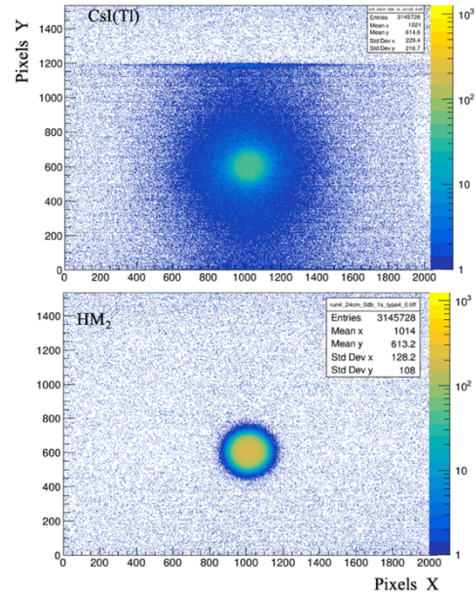

Figure 2: Image of a β$^-$ from $^{90}$Sr on 1.25 mm thick CsI(Tl) (top) and 0.435 mm $HM_2$ (bottom).

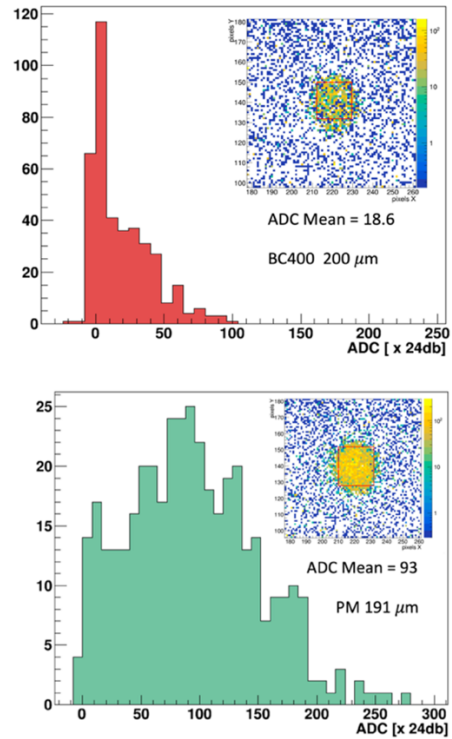

Figure 3: Inset is image of $^{90}$Sr β$^-$ source on BC-400 (top) and PM (bottom). ADC spectra are from the red frame signal regions.

*FRIB Beam Test Results*

At FRIB PM and $HM_2$ scintillators were tested. (Here HM=type 2). In-house diagnostic systems were used as a beam current reference, including a Faraday cup above 25,000 particles/s (pps), a silicon detector for counting rates below 100 pps, and a microchannel plate (MCP) that was used for the $10^{2-3}$ pps range. Intermediate hit rates (> $10^3$ pps) were determined by calibrated beam attenuators. The SBM was evaluated at rates that extend from 520,000 pps down to effectively single, isolated particles at a total rate of ~5-10 pps distributed over an area of several $mm^2$. Figure 4 specifically shows these single particle hits on HM scintillator for a 1s frame exposure (1 Hz frame rate). Figure 5 shows the hit profile for a ~50 pps beam for HM (left) and a reference CsI(Tl) 1.25 mm thick, unpolished scintillator (right) for comparison. In CsI(Tl) the beam is barely visible while in HM a clear beam image emerges.

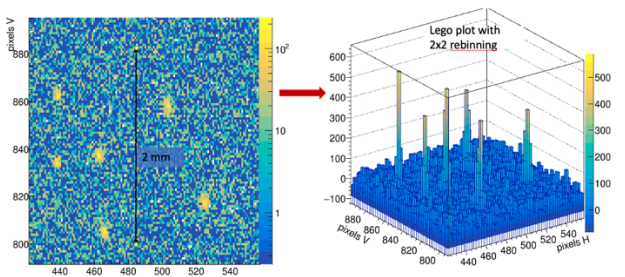

Figure 4: Individual $^{86}$Kr hits on HM type scintillator. Camera frame exposure time = 1 s, beam rate ~5-10 pps.

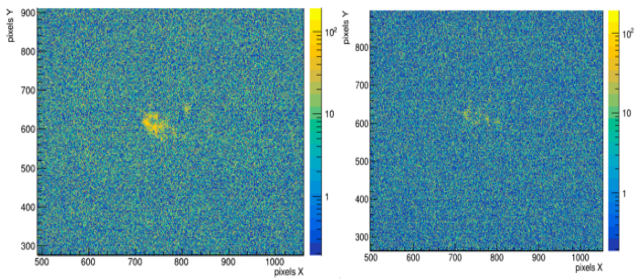

Figure 5: Image of a 50 pps beam on HM scintillator (left) and CsI(Tl) (right).

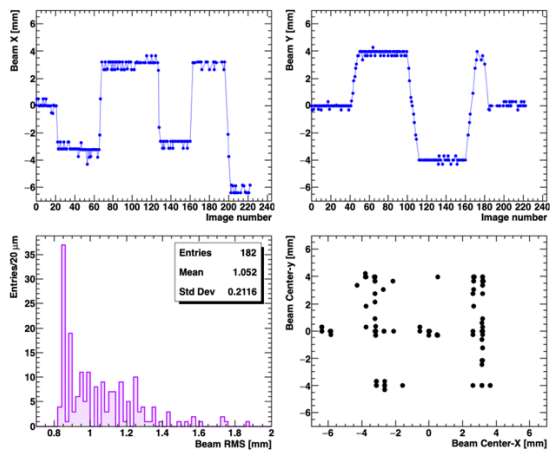

Figure 6: Tracking history of a low-rate (50 pps) beam motion. Top left, right: X and Y coordinate history where image frames are 1 s apart. Bottom left is the RMS beam radius and bottom right shows the 2D map of beam centroid positions.

The integrated photon signal over the beam area is much larger for HM vs CsI(Tl).

An updating low rate beam position tracking validation was conducted in real-time by the control room engineers incrementally translating the above 50 pps beam over a ~1 cm square region. The SBM tracking algorithm finds the beam centroid in the background subtracted images, performs perspective transforms and the necessary rotations to then present the centroids and widths in the beam coordinate system to the control room *in real-time*. Figure 6 (top left, top right) shows the history of the motion along the beam X and Y coordinates; the bottom left shows the RMS beam width and the bottom right shows the full 2D reconstructed beam position movement over a period of a few minutes.

**Linearity and dynamic range:** The signal response vs beam rate of HM scintillator at FRIB was measured by integrating the background-subtracted ADC signal over the entire area of the beam profile. This integral was then normalized to the **average** single particle signal. To obtain the single particle normalization constant, each of several hundred isolated $^{86}$Kr hits in all low-rate data sets, as exemplified in Figure 4, were integrated out to the maximum hit radius. The results of this measurement are shown in Figure 7. The X coordinate axis reports the corresponding beam currents provided by FRIB beam control. These are based on the outputs of three rate-specific FRIB detectors (silicon strip detector, MCP, Faraday cup), and on the use of calibrated beam attenuators that provide current estimates. The estimated total rate uncertainty of these rates is ~20%. The salient feature of Figure 7 is that the SBM result is linear within the fit error, and it tracks the beam current from single hits to five orders of magnitude.

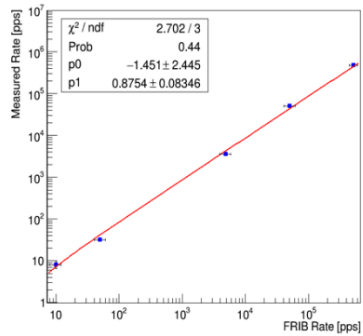

Figure 7: The SBM measured beam rate in HM scintillator plotted against the rates provided from three different FRIB reference detectors.

*MIBL Results*

The SBM was staged in a higher current (1-10nA), low energy (1-5.4 MeV) proton beam with the objective of measuring radiation degradation in PM type scintillators at high ionizing dose rates. In these experiments the beam was rastered over a 1 cm$^2$ aperture to produce a current density at the scintillator target of 4 nA/cm$^2$. Thin (1-200 μm) PM scintillators were evaluated for signal output, image quality and radiation degradation. For these high exposure rates in vaccuum of over 300 Gy/s, the average signal was observed to diminish by 0.5 ± 0.1%/kGy as shown in Figure 8, for PM samples of varying thickness. Notably, when the samples are left in air, they tend to recover over a time of a few hours. Figure 9 shows the UV excited scintillation light yields measured post-irradiation for several PM samples of varying thickness, and for 20 kGy integrated dose. For reference, a commercial scintillator type EJ260 is also shown. The commercial product suffers greater signal loss, but recovers partially in 2 hours (later data were not acquired). Other measursments of EJ260 show, after 2 months, a 10% drop in transmittance at 550 nm for 5x10$^{11}$ protons/cm$^2$ at 150 MeV (approximately 43 kGy) [5].

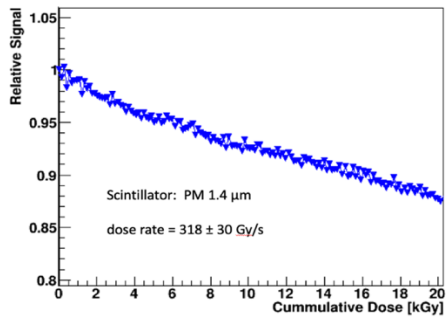

Figure 8: Relative signal in 1.4 μm thick PM scintillator during exposure at MIBL from 5.4 MeV protons at a dose rate of 318 Gy/s.

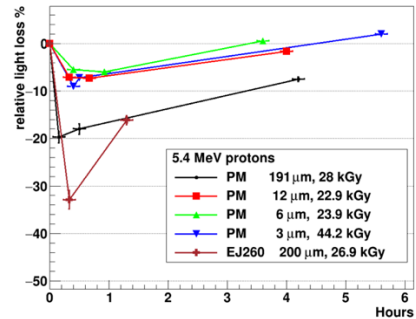

Figure 9: The UV induced scintillation signal strength vs recovery time in air for PM scintillators after exposure to 5.4 MeV protons.

## CONCLUSIONS

The SBM is a verstaile instrument with a demonstrated ability to profile ion beams from low rates (single particles) to 10 nA currents using two novel scintillators mounted on a mult-target casette. High sensitivity HM scintillators readily detect single particles and low rate beams. PM type scintillators are shown to produce stronger signals than common commercial varieties. The PM can also be ultra-thin, with micron thickness and be partially or fully transmissive to ion beams. Scintillators are inserted into/out of the beam without breaking vacuum. A data acquistion and analysis software package is able to detect, measure and display beam profiles in real-time, enabling the SBM to be an effective beam tuning instrument.